\RequirePackage{fix-cm}
\documentclass[smallextended]{svjour3}       
\smartqed  
\usepackage{graphicx}
\usepackage[round,authoryear]{natbib}
\PassOptionsToPackage{hyphens}{url}\usepackage{hyperref}
\usepackage[toc,page]{appendix}
\usepackage{float}
\usepackage{enumitem}
\usepackage{subfig}
\usepackage{multirow}
\begin{document}

\title{On the interplay of data and cognitive bias in crisis information management
}
\subtitle{An exploratory study on epidemic response}


\author{David Paulus        \and
        Ramian Fathi \and
        Frank Fiedrich \and
        Bartel Van de Walle \and
        Tina Comes 
}


\institute{D. Paulus (corresponding author) \at
              \email{d.paulus@tudelft.nl}             
           \and
           R. Fathi \at
              fathi@uni-wuppertal.de
                   \and
           F. Fiedrich \at
              fiedrich@uni-wuppertal.de
                   \and
           B. Van de Walle \at
              vandewalle@merit.unu.edu
              \and
              T. Comes \at
              T.Comes@tudelft.nl
}

\date{Received: date / Accepted: date}

\maketitle

\begin{abstract}Humanitarian crises, such as the 2014 West Africa Ebola epidemic, challenge information management and thereby threaten the digital resilience of the responding organizations. Crisis information management (CIM) is characterised by the urgency to respond despite the uncertainty of the situation. Coupled with high stakes, limited resources and a high cognitive load, crises are prone to induce biases in the data and the cognitive processes of analysts and decision-makers. When biases remain undetected and untreated in CIM, they may lead to decisions based on biased information, increasing the risk of an inefficient response. Literature suggests that crisis response needs to address the initial uncertainty and possible biases by adapting to new and better information as it becomes available. However, we know little about whether adaptive approaches mitigate the interplay of data and cognitive biases. 

We investigated this question in an exploratory, three-stage experiment on epidemic response. Our participants were experienced practitioners in the fields of crisis decision-making and information analysis. We found that analysts fail to successfully debias data, even when biases are detected, and that this failure can be attributed to undervaluing debiasing efforts in favor of rapid results. This failure leads to the development of biased information products that are conveyed to decision-makers, who consequently make decisions based on biased information. Confirmation bias reinforces the reliance on conclusions reached with biased data, leading to a vicious cycle, in which biased assumptions remain uncorrected. We suggest mindful debiasing as a possible counter-strategy against these bias effects in CIM.

\end{abstract}

\section{Introduction}
\label{intro}
Infectious disease outbreaks have been on the rise \citep{Smith2014}, with the COVID-19 pandemic being the prime example that epidemics, if not controlled, lead to severe humanitarian crises and exacerbate poverty and hunger in the Global South \citep{UnitedNations2021b}. To respond to epidemic crises, information is central. Previous research has advocated for digital resilience via information systems, models, and algorithms that address the deluge of information and foster the stability of the digital ecosystem itself \citep{schemmer2021conceptualizing}. \cite{Constantinides2020} define digital resilience as \textit{"[...] the phenomena of designing, deploying, and using information systems to quickly recover from or adjust to major disruptions from [...] shocks."} Crises, however, put digital resilience to the test, especially the ability to rapidly adapt to a dynamic and highly volatile information environment.

The exceptional circumstances of crises put enormous pressure on crisis information management (CIM) as it needs to happen rapidly, despite tremendous uncertainty, and is often heavily resource-constrained \citep{Schippers2020, Comes2020}. These characteristics pose a double challenge: (a) data may not be available or is biased given limited access or data collection regimes, or it may be noisy, uncertain, and conflicting; and (b) the cognitive processes of crisis information managers and decision-makers may be under strain, given the urgency and high stakes of the situation. 

Regarding (a), in crises, relevant data is often unavailable because of access constraints or destruction of infrastructure or because decisions have to be made quicker than it takes to collect and analyze data \citep{Fast2017}. This can lead to representational bias in data that potentially over- or under-represent issues, social groups, or geographic areas \citep{Fast2017, Galaitsi2021}. If such biases remain undetected and untreated in CIM, information products used to support decision-making will also become biased. 

Regarding (b), crises pose significant challenges to the cognitive processes of information managers and decision-makers. Humans tend to be influenced by cognitive biases, especially in situations of urgency, uncertainty, risks, and high-stakes \citep{Phillips-Wren2019}. The concept of cognitive biases originates from the idea of bounded rationality that postulates, human thinking (within the complex world surrounding it) is limited, which prevents people from being purely rational \citep{Simon1955}. Confirmation bias is among the most prominent cognitive biases in crises \citep{Brooks2020, Comes201656, Modgil2021}. It leads people to search and select information that confirms their previous assumptions and decisions and neglect disconfirming information \citep{Nickerson1998}. Consequently, crisis responders might disregard valid and important information only because it conflicts with or does not confirm their initial assumptions. 

We argue that the interplay of data bias and confirmation bias threatens the digital resilience of crisis response organizations. The consequences for crisis response can be particularly severe when data bias and cognitive bias reinforce each other in sequential decisions over time. When initial assumptions are made based on biased data, confirmation bias may lead people to further rely on information that confirms their initial biased assumptions. This might lead to a vicious circle that hampers adaptation and prolongs initially wrong decisions rather than correcting them. Conventionally, the literature suggests that decisions in crises need to be adaptive to new information \citep{turoff2004design}. The principle of strengthening the adaptive capacity to manage uncertainty is underlying a broad range of literature on adaptive management in crises and (digital) resilience \citep{tim2021digital, schiffling2020implications}. However, we know little about the effectiveness of such adaptive approaches against the backdrop of combined data and confirmation bias. 

A potential counter-strategy to mitigate the negative consequences of biases on CIM is mindful debiasing. Mindfulness means being more aware of the context and content of the information one is engaging with \citep{Langer1992}, thereby becoming less prone to confirmation bias \citep{Croskerry2013}. In a mindful state, information managers are more open to new and different information \citep{Thatcher2018}. In contrast, when being less mindful, people rely on previously constructed categories and neglect the potential novelty and difference within newly received information \citep{Butler2006}.

This exploratory study investigates the interplay of data and confirmation bias in a sequential setup. Through a three-stage experiment with experienced practitioners, we studied how our participants dealt with biased data, and in how far they were able to correct initial decisions, or whether path-dependencies to biased decisions emerged. Based on our findings, we outline how mindful debiasing can support the detection and mitigation of data and confirmation biases in crisis response.  

The remainder of this paper is structured as follows: the next section reviews the relevant literature related to CIM, digital resilience and biases, and provides the research gap and research questions this paper is addressing. Section 3 describes the research design and methods, and Section 4 provides the results from our experiment. In Section 5, we discuss our contributions to literature and practice as well as future research avenues. In Section 6, we reflect on the limitations of this exploratory study, and Section 7 concludes the paper.

\section{Background}

\subsection{Crisis information management}

\subsubsection{Approaches and tools to crisis information management}
Crisis information management (CIM) entails the formulation of data needs, identification of data sources, data collection, cleaning and structuring, data analysis, and the design and development of information products \citep{Currion2007}. The objective of CIM is to support decision-making by providing trustworthy, accurate, and actionable information. With the rise of Big Data and Artificial Intelligence, larger humanitarian organizations have invested in analytics capacity \citep{Akter2019}. While the potential for working with unstructured data for predictive analytics has been recognized, many humanitarian organizations active in the Global South do not possess the resources for large investments into information technology and statistical sophistication \citep{prasad2018big, baharmand2021developing}. In these contexts, large parts of CIM are still supported through common office information systems such as Microsoft Excel and Google Spreadsheets \citep{UnitedNations2020}. These are used, amongst others, to store survey responses, conduct data integration, and develop information products, e.g., maps, tables, and infographics \citep{Thom2015}.

\begin{figure}[b]
\caption{Map comparison for Sierra Leone during the 2014-2016 Ebola outbreak.}
\label{fig:exampleinformationproduct}
\centering
    \subfloat[\centering MapAction map showing mobile phone coverage]{{\includegraphics[width=5.5cm]{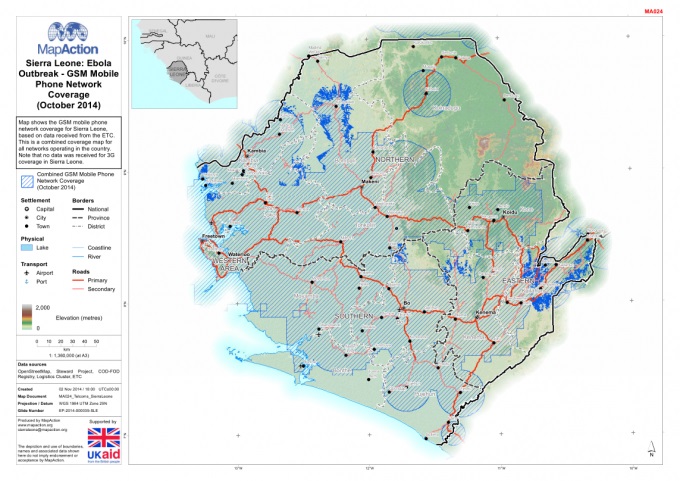} }}%
    \qquad
    \subfloat[\centering WHO map of hot-spots of Ebola cases. ]{{\includegraphics[width=5.5cm]{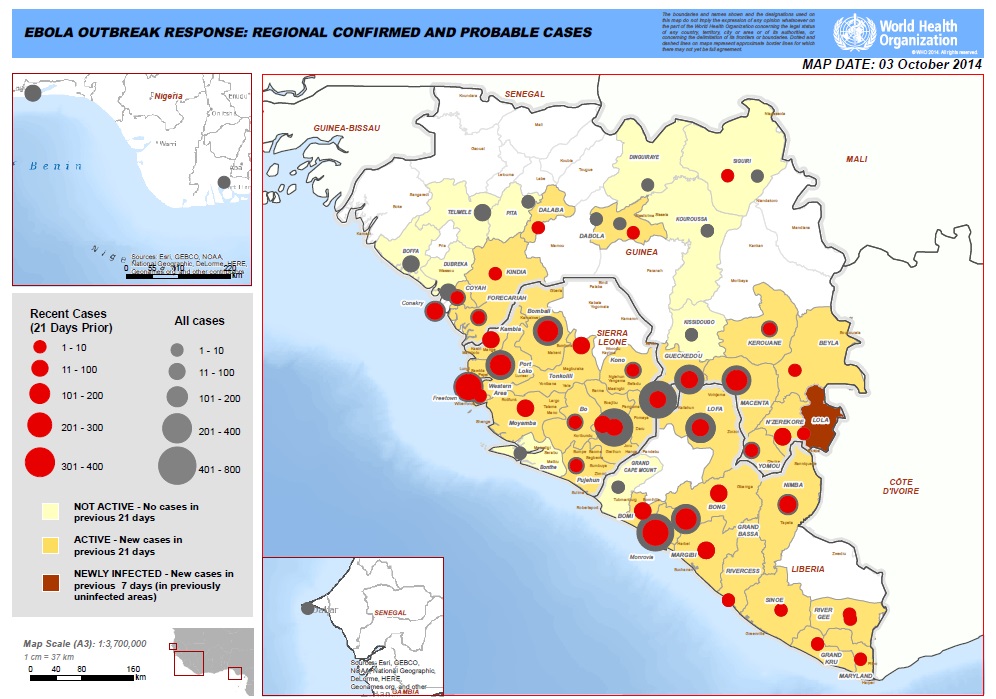} }}%
\end{figure}

Especially in sudden-onset disasters, organizations frequently surge additional data analyst capacity to rapidly strengthen their CIM and digital resilience. Often, these are remotely working \textit{digital volunteers}, that have been regarded as cost-effective, additional analyst capacities to support CIM \citep{Poblet2018, Castillo2016}. These external analysts contribute to CIM by supporting tasks such as data collection, analysis as well as the development of information products for decision support \citep{Chaudhuri2020, Hughes2015, Karlsrud2017}. External analysts have also contributed to epdiemics CIM, e.g., in the 2014 West Africa Ebola outbreak \citep{Hellmann2016}, or the ongoing Covid-19 response \citep{Fathi2021}. 
 
Figure~\ref{fig:exampleinformationproduct} shows on the left side an information product developed by external analysts during the 2014 Ebola outbreak. The product highlights the major challenges of access to data and shows that the mobile phone network corresponds to the areas of the officially reported cases (WHO map at the right-hand side of Figure~\ref{fig:exampleinformationproduct}), clearly an indication of the widespread data biases, whereby access and phone coverage hampered reporting. Other information products created through such joint CIM processes include Excel and Google spreadsheets, graphs, and 1-pager summarizing results of social media data analyses \citep{Hughes2015}.

Tasks and responsibilities frequently shift in crises \citep{nespeca2020towards}, requiring information managers and decision-makers to interact with data in different ways. While external analysts are primarily turning raw data into information, decision-makers are concerned with interpreting the situation and putting received information into context by using experience, communicating with partners, acting, and reacting.

\subsubsection{Sensemaking and situational awareness}
While much work on decision-making in crises focuses on optimizing for isolated decisions, crises are typically characterized by nested and interdependent decisions, driven by cognition and experience. This process is recognized by the literature on \textit{sensemaking}, whereby decisions are part of a broader collective process of meaning-making \citep{weick1995sensemaking, Klein2006, Comes2020}. Important components of sensemaking are information seeking, processing, creating, and using \citep{muhren2008sensemaking}. Data-driven approaches, e.g., predictive analytics, can support sensemaking by revealing internal and external cues. Sensemaking is also influenced by an organization's mandate, strategy and modes of operation \citep{Zamani2021}, and especially describes how people deal with 'gappy' information environments \citep{muhren2008sensemaking}.

Early studies on the work of external analysts emphasized the added value they bring to CIM by their remote and flexible structures \citep{Meier2012, Ziemke2012, Bott2012}. It has been argued that their work contributes to the situational awareness of response organizations \citep{Hughes2015, Starbird2011a}. To achieve situational awareness successfully, however, it is important to switch between goal-driven and data-driven approaches \citep{Endsley1995, R.Endsley2003, Fromm2021}. While for goal-driven approaches, informational cues are intentionally considered in the pursuit of a set goal, data-driven approaches refer to open exploration of perceived cues that can lead to changes in priorities and readjustments. Situational awareness requires to alternate between these two forms because stringent goal-focus will lead to neglect of cues in the data, while stringent data-focus will be perceived as overly taxing \citep{Fromm2021}.

\subsection{Digital resilience and crisis information management}

\subsubsection{Defining digital resilience}
There are diverging perspectives on what constitutes digital resilience and whether it plays at the level of the physical infrastructure, the people or groups using the infrastructure, or the interplay among both. Some authors focus on the impact of digital technology on the user, stressing the importance of (access to) information in crises. For instance, according to \cite{Wright2016}, \textit{"digital resilience means that to the greatest extent possible, data and tools should be freely accessible, interchangeable, operational, of high quality, and up-to-date so that they can help give rise to the resilience of communities or other entities using them."} Others focus on the resilience capabilities of individuals to process digital data and engage with virtual environments \citep{UKCouncilforInternetSafetyUKCIS2019A}. 

[Here, we take an information systems perspective, understanding digital resilience as a phenomenon that emerges from the interaction of people with data through digital tools and infrastructure. We follow a crisis-related definition that describes digital resilience as a means to cope with disruptions: \textit{"[...] digital resilience [...] refer[s] to the phenomena of designing, deploying, and using information systems to quickly recover from or adjust to major disruptions from [...] shocks."} \citep{Constantinides2020}. Crisis information management needs to foster digital resilience by supporting flexibility, agility, and adaptability \citep{turoff2004design}. Our definition also covers specific aspects of digital resilience during epidemics \citep{Marifat2020}, namely the collection and analysis of outbreak data, as well as the use of analysis results to inform crisis response. Since CIM incorporates data collection, analysis, and sharing to support crisis decisions, it is directly linked to digital resilience.

\subsubsection{Challenges to digital resilience in crisis information management}\label{sec:CIMbias}

Previous literature identified several challenges to CIM that affect different functions \citep{VandeWalle2015, lauras2015event} at different hierarchical levels \citep{bharosa2010challenges}. We argue that data and cognitive biases can emerge as consequences to these challenges and affect CIM by posing threats to digital resilience in terms of hampering the rapid recovery from crises. We use the challenges described below to design our experiments, described in Section 3.

Information has to feed into the fast crisis decision-making process \citep{warnier2020humanitarian, lauras2015event, turoff2004design}. The time pressure reinforces the tendency to focus only on information that is immediately available \citep{higgins2013improving}, which may induce a range of biases \citep{maule2000effects}. Information needs also rapidly change during different crisis stages \citep{Hagar2011, Gralla2015, nespeca2020towards}, posing challenges to the agility and flexibility of information management \citep{lauras2015event}. 

As the destruction of infrastructure or lack of access may affect different regions to different degrees \citep{altay2014challenges}, datasets are often geographically imbalanced or biased. Demographic biases can influence the data further. Especially in the Global South, the most vulnerable groups might not have access to mobile phones and therefore are not included in mobile phone data to track and trace population movements \citep{IOM2021}. Under-representation of geographic areas or social groups can lead to violations of the humanitarian imperative to '\textit{leave no one behind}' \citep{VandeWalle2015}. 

Relevant information about the crisis situation is often uncertain. Uncertainty is an umbrella term for information that is unavailable, incomplete, ambiguous, or conflicting \citep{COMES2011108, Tran2021}. To reduce uncertainty, people likely use the tools and methods they are most familiar with. This behavior could lead to what is known as the law-of-the-instrument, which states that people tend to overly rely on a particular familiar tool \citep{Johnson2020}.

The high volume, velocity, and variety of irrelevant data can quickly lead to information overload, particularly when the veracity of data has to be evaluated as well \citep{Schulz2012}. This issue has become particularly prominent with the ubiquity of social media \citep{gupta2019big}, which makes it virtually impossible to filter and process all available data on time \citep{Starbird2011, van2016improving}. Information overload has been shown to induce confirmation bias \citep{goette2020information}. Confronted with an overload of information, it is hard to identify any gaps in the available data, leading to exploiting what is known rather than exploring what could be known \citep{Comes2020}. 

In the high stakes decision contexts of humanitarian crises, tremendous potential losses are combined with the irreversibility of decisions \citep{kunreuther2002high}. High stake situations have been shown to induce a large number of biases, ranging from a tendency to focus on short-term perspectives as well as an over-reliance on social norms and emotional cues \citep{kunreuther2002high}. For example, high-stakes decisions can lead decision-makers to exert groupthink, which is manifested by overconfidence and a strive for in-group harmony, rather than critical self-reflection \citep{Kouzmin2008}.

\subsection{Biases in crisis information management}
As we have shown, the characteristics of crises provide a breeding ground for data biases and cognitive biases \citep{Comes201656}. Here, we zoom into two of the most prominent biases that are relevant in the interplay of information and decision-making: data and confirmation bias. 

\subsubsection{Data bias in crisis information management}
Data can become biased due to historical, social, political, technical, individual, and organizational reasons \citep{Jo2020}. Representational data bias is among the most common forms and a broad category of data bias. It comes from the \textit{"divergence between the true distribution and digitized input space"} (ibid.). In practice, that often means that a dataset systematically deviates from the real-world phenomenon the data is supposed to represent, for example, leading to the under-representation of geographic areas or social groups.   

Data bias can be understood as a flaw of a dataset, negatively affecting the quality of the data and potentially causing damages and losses in organizational processes \citep{Storey2012}. Especially in sensitive contexts, data bias has been shown to replicate and reinforce existing inequalities \citep{Jacobsen2019, Bender2020}. Urgency and overload combined with uncertainty are common causes for data bias in crises \citep{Fast2017}. 

In epidemic response, the misrepresentation of infection rates has been documented during the 2014-2016 Ebola outbreak in West Africa \citep{Fast2017}. Similarly, during the COVID-19 pandemic, different testing, tracing, or counting strategies have resulted in incomplete datasets and incomparable statistics \citep{Fenton2020}. 

We look at representational bias in two key variables for epidemic response: numbers of infections and treatment capacity. Representational bias in those two variables can lead to a flawed understanding of the outbreak's severity and the available capacity, leading to misallocations and delayed or ineffective response.

One of the hopes in using additional analytic capacity is that this additional capacity identifies additional information and thereby helps overcome data bias. To test if additional external capacity actually helps in overcoming data bias, we draw inspiration from traditional hidden profile experiments \citep{Stasser1985, Lightle2009}. These experiments evaluated groups' decision-making performance. Group members received two sets of information, one set that contains the same information for all group members and another set that is different between group members. Only by joining the different, individual information sets together groups can identify the hidden profile, which is crucial to make the optimal decision. Hidden profile experiments have shown that generally groups overly discuss common information and neglect individual information so that the hidden profile remains hidden and the groups make an inferior decision \citep{Stasser1985, Lightle2009}. This behavior was also found in experiments on crisis decision-making \citep{Muhren2010}. However, previous experiments did not specifically look at representational bias in crises and whether adaptive approaches to surge additional analyst capacities help to improve the identification and mitigation of biases.

\subsubsection{Confirmation bias in crisis information management}
\label{lab:confbiaslit}
A cognitive bias that hampers adequate adaptation to new information is \textit{confirmation bias}. Research on confirmation bias has shown that people tend to limit their information retrieval efforts to information that is more likely to confirm their assumptions \citep{Nickerson1998}. Because information that opposes preliminary assumptions increases discomfort \citep{Hart2009}, it may be discarded, and wrong assumptions remain undetected, leading to flawed decision-making \citep{NAP19017}. Confirmation bias, like cognitive biases in general, are often characterized as a byproduct of information processing limitations: because of urgency and overload, people use biases as mental shortcuts to judge and decide quickly. 

The urgency of crises likely fosters confirmation bias because relying on already formed assumptions accelerates decision-making. Domain experts, however, can show the opposite behavior and deliberately seek disconfirming information \citep{Klein2006}. Counterfactual mindsets have been shown to be an effective debiasing strategy \citep{kray2003debiasing}. However, we know little about the potential influence of confirmation bias on the information search and selection behavior of experienced crisis responders.

In this study, we investigate if crisis decision-makers and analysts are susceptible to confirmation bias and if they search for non-confirmatory data as a debiasing strategy. It could be possible that the deliberations between experts induce counterfactual mindsets, which, in turn, lead to a more critical assessment of prior decisions. However, path-dependencies may arise, whereby confirmation bias leads decision-makers and analysts to confirm assumptions in subsequent decisions, even though they were made based on biased data.

Previous research measured confirmation bias through tasks with two parts \citep{Jonas2001, Fischer2011}. First, participants made a preliminary decision between two options on a certain matter. Then, they were presented a set of information, which often are summaries of articles on the matter participants just made their preliminary decision on. For example, ten summaries of articles are presented, five supporting participants' preliminary choice, and five opposing it. Participants are then asked to select the articles they would like to receive in full. The experiment finishes, and participants are told there will be no full articles because it is unnecessary for the experiment. The researcher later counts the numbers of selected supporting and opposing article summaries and conducts a significant test for the difference. If significantly more supporting summaries were selected, we  speak of confirmation bias. 

\subsection{Research gap and research questions}
In dynamic situations such as crises, information on the best course of action continuously changes. Therefore, the literature advocates for agile and adaptive management in epidemics \citep{Merl2009, Janssen2020} or, more generally, in crises \citep{Charles2010, Anson2017, schiffling2020implications, turoff2004design}. 

Response organizations often lack sufficient capacities to respond. Therefore, remotely working external analysts are added as surge capacity. There is some hope that via this additional capacity, exploratory search strategies may be favored that help overcome the responsive and exploitative strategies of decision-makers. At the same time, the remote nature of the work of analysts may add to the biases they are subject to \citep{Comes201656} and may make especially data interpretation harder \citep{comes2016information}. Therefore, it is not yet known how and in how far the interplay of analysts and decision-makers in sequential decisions reduces or amplifies biases. In this paper, we investigate whether the surge of additional analyst capacity is effective to mitigate bias effects.

In sequential decisions, initial biases might limit the ability to effectively adapt, even though adaptation is widely described in the crisis management literature as key to managing the uncertainties and data biases that often prevail at the onset of a crisis \citep{mendonca2001decision, quarantelli1988disaster}. Potentially, representational data bias and confirmation bias reinforce each other, leading to amplified biases. This is especially harmful if path-dependencies arise whereby the initial data bias does not only influence initial decisions but leads to flawed decision trajectories through confirmation bias.

\begin{figure}[h]
\caption{External analyst-supported crisis information management process. Source: authors.}
\label{fig:crisisinformationmanagement}
\centering
\includegraphics[width=\textwidth]{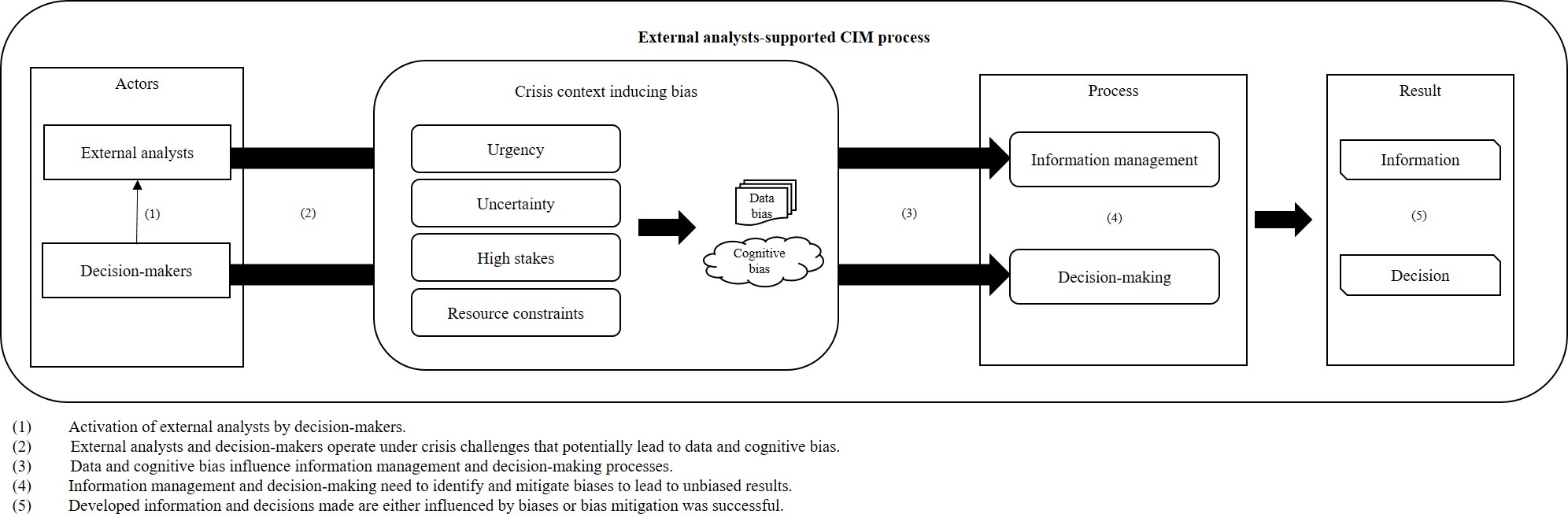}
\end{figure}

Figure \ref{fig:crisisinformationmanagement} depicts the interaction of the identified main challenges within the external analyst-supported CIM process. The response organizations activate external analysts in the first step (1). In steps (2) and (3) external analysts and decision-makers conduct information management and decision-making under the influence of the crisis, which can lead to biases. Information management and decision-making need to identify and mitigate biases to lead to unbiased results (4). Finally, the resulting information and decision are either influenced by biases, or bias mitigation was successful (5).

We are interested in (RQ 1) whether the surge of external analysts leads to unbiased information products for decision support, (RQ 2) if the joint CIM process between analysts and decision-makers facilitates debiasing, and (RQ 3) if data bias and confirmation bias reinforce each other leading to path dependencies in sequential decisions. We address the following research questions:

\begin{enumerate}[align=left]
\item[RQ 1:] Is surging external analysis capacity effective in identifying and mitigating data bias?
\item[RQ 2:] How do external analysts and decision-makers jointly handle data bias in the decision process?
\item[RQ 3:] Does confirmation bias create path dependencies whereby biased assumptions persist in sequential decisions?

\end{enumerate}

We used an exploratory, three-stage experiment to examine these research questions, which is described in detail in the next section.

\section{Research Design \& Methods}
\label{sec:method}
\begin{figure}[h]
\caption{Research design.} 
\vspace*{+2mm}
\label{fig:researchdesign}
\centering
\includegraphics[width=300pt]{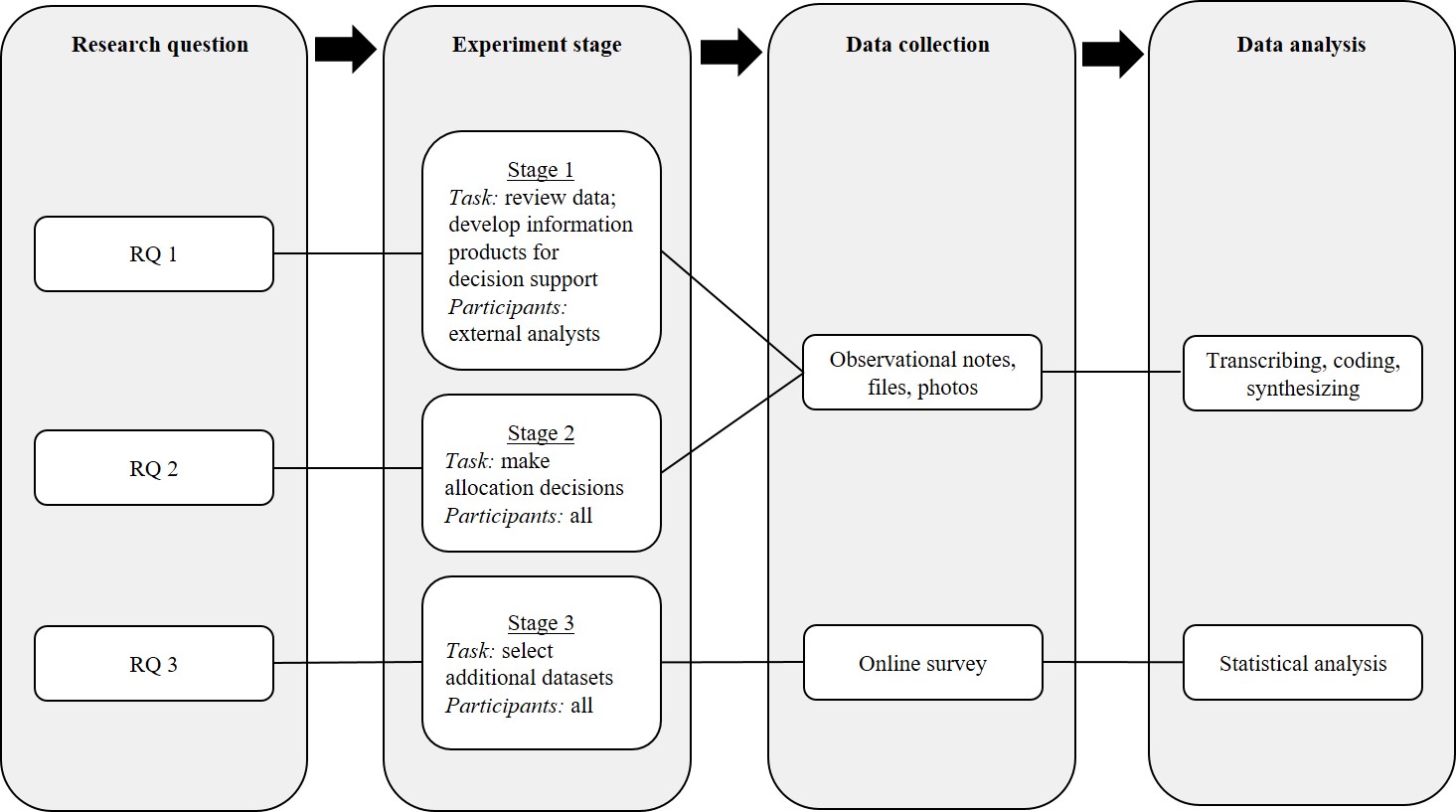}
\end{figure}

We conducted an exploratory study with three stages to address the three research questions (Figure \ref{fig:researchdesign}). RQ 1 and RQ 2 were addressed through a scenario-based workshop with experienced practitioners in the fields of crisis decision-making and external analysis for CIM support. RQ 3 was addressed through an online survey with the same participants. Figure~\ref{fig:researchdesign} depicts the research questions together with the corresponding experiment stages, data collection, and analysis methods.

The experiment was designed to observe the crisis information management and decision-making process in a controlled environment. The controlled environment enables observation without interfering with the real response and allows us to conduct the experiment with three different groups. Yet, by designing realistic information flows, creating time pressure and providing the typical tools, the scenario is sufficiently realistic enough to inspire the same ways of thinking that external analysts or decision-makers also show in real epidemics. Through this setting, it was possible to observe the practices, communication and interactions within and between the participant groups. The experiment took place at the TU Delft Campus in The Hague in January 2020.

\subsection{Participants}
\subsubsection{Recruitment}
Participants had to have work experience as external analysts or decision-makers in crises to be eligible for participation. The recruitment was done based on the competencies required to fulfill the tasks of our experiment. These competencies included technical skills such as merging tabular data in MS Excel or a similar tool and developing and interpreting crisis information products such as maps and graphs. In addition, participants needed to be affiliated to an established crisis response organization, data analytics organization, or research institute on crisis or epidemic management. The authors had contacts to a network of potential candidates through previous research. This enabled us to recruit participants who had the required skills and experience. The participants were recruited internationally from various countries. Table \ref{tab:participants} lists the descriptive information of our participants.

\begin{table}[]
\caption{Descriptive information of all participants during the experiment. EA = External analyst, DM = Decision-maker.}
\label{tab:participants}
\resizebox{\textwidth}{!}{%
\begin{tabular}{llll}
\hline
\multicolumn{1}{c}{\textbf{Group}} & \multicolumn{1}{c}{\textbf{Role}} & \multicolumn{1}{c}{\textbf{Organization}}                                                                     & \multicolumn{1}{c}{\textbf{Competencies}}                                                                                                                         \\ \hline
Reloupe                            & EA                                & \begin{tabular}[c]{@{}l@{}}Humanitarian Openstreetmap \\ Team\end{tabular}                              & \begin{tabular}[c]{@{}l@{}}Mapping and open data for \\ humanitarian action\end{tabular}                                                                          \\
                                   & EA                                & MapAction                                                                                                     & \begin{tabular}[c]{@{}l@{}}Mapping and open data for \\ humanitarian action\end{tabular}                                                                          \\
                                   & EA                                & Mark Labs                                                                                                     & \begin{tabular}[c]{@{}l@{}}Data analytics for environmental and \\ social transformation\end{tabular}                                                             \\
                                   & EA                                & 510 (Red Cross)                                                                                               & \begin{tabular}[c]{@{}l@{}}Emergency data support, predictive \\ impact analysis and digital risk assessment\end{tabular}                                         \\
                                   & DM                                & TU Delft                                                                                                      & Student with no prior experience                                                                                                                                  \\
                                   & DM                                & Red Cross                                                                                                & \begin{tabular}[c]{@{}l@{}}Emergency response, volunteer   \\ assistance, emergency training\end{tabular}                                                         \\
                                   & DM                                & Dorcas                                                                                                        & Poverty reduction and crisis response                                                                                                                             \\
                                   & DM                                & US Department of State                                                                                        & Senior humanitarian analyst                                                                                                                                       \\
                                   & DM                                & Municipal Health Service                                                                                      & Doctor of infectious disease control                                                                                                                              \\
Republic                           & EA                                & Standby Taskforce                                                                                             & \begin{tabular}[c]{@{}l@{}}Mapping and open data for   \\ humanitarian action\end{tabular}                                                                        \\
                                   & EA                                & \begin{tabular}[c]{@{}l@{}}Virtual Operations Support \\ Team \end{tabular}                             & Social media data analyis in crisis response                                                                                                                      \\
                                   & EA                                & Standby Taskforce                                                                                             & \begin{tabular}[c]{@{}l@{}}Mapping and open data for   \\ humanitarian action\end{tabular}                                                                        \\
                                   & EA                                & TU Delft                                                                                                      & Student with no prior experience                                                                                                                                  \\
                                   & DM                                & ZOA                                                                                                           & \begin{tabular}[c]{@{}l@{}}Emergency relief and reconstruction \\ of regions struck by disasters or conflicts\end{tabular}                                        \\
                                   & DM                                & Red Cross                                                                                                & Emergency Relief Coordinator                                                                                                                                      \\
                                   & DM                                & World Vision                                                                                                  & \begin{tabular}[c]{@{}l@{}}Disaster management, economic   \\ development, education, faith and \\ development, health and nutrition and   \\ water.\end{tabular} \\
Noruwi                             & EA                                & 510 (Red Cross)                                                                                               & \begin{tabular}[c]{@{}l@{}}Emergency data support, predictive \\ impact analysis and digital risk assessment\end{tabular}                                         \\
                                   & EA                                & MapAction                                                                                                     & \begin{tabular}[c]{@{}l@{}}Mapping and open data for humanitarian \\ action\end{tabular}                                                                          \\
                                   & EA                                & Leiden University                                                                                             & \begin{tabular}[c]{@{}l@{}}Development of data-driven decision support\\ tools for humanitarian organizations\end{tabular}                                        \\
                                   & EA                                & TU Delft                                                                                                      & Student with no prior experience                                                                                                                                  \\
                                   & EA                                & Humanity Road                                                                                                 & Social media data analyis in crisis response                                                                                                                      \\
                                   & DM                                & \begin{tabular}[c]{@{}l@{}}TU Delft / European Commission\\ Humanitarian Aid Office\end{tabular}              & Emergency and crisis management                                                                                                                                   \\
                                   & DM                                & \begin{tabular}[c]{@{}l@{}}Maastricht University Faculty of\\ Health, Medicine and Life Sciences\end{tabular} & Public health expert                                                                                                                                              \\
                                   & DM                                & Ministry of Foreign Affairs (NL)                                                                              & Senior humanitarian advisor                                                                                                                                       \\ \hline
\end{tabular}%
}
\end{table}

\subsubsection{Sample}
Twenty-four participants participated in the experiment, of which twenty-one were experienced in crisis management (eleven external analysts and ten decision-makers), and three were students. We added three students to create an element of reality to the group compositions as staff turnover is high in crisis response teams, with new and inexperienced staff needing to be integrated \citep{St.Denis2012, FATHI2020102174}. Based on the background and experience of the participants, they were given either the role as an external analyst or as a decision-maker. Participants within the group of external analysts were part of professional disaster relief organizations as well as organizations representing different fields of expertise such as digital mapping, social media analysis, and data analytics. The group of decision-makers consisted of representatives from different governmental and non-governmental crisis response organizations from numerous countries, including The Netherlands, Germany, the United Kingdom, and the United States. Table \ref{tab:participants} gives an overview of all participants, their corresponding organizations, and competencies.

Recruiting experienced professionals for a scientific experiment leads to a smaller pool and thereby also lower participant numbers as compared to experiments with students or the general public. As the objective of this exploratory experiment was to gain insights into information management and decision-making approaches by actual practitioners, relying on samples drawn from student populations or the general public would have been inadequate.  

Our sample size is in a similar range as comparable exploratory studies on information systems and information management \citep{Antunes2020}. Such exploratory studies provide a valid approach to build theory and identify metrics, mechanisms, processes, and concepts that can be investigated further in subsequent empirical research \citep{Antunes2020}.

\subsubsection{Group compositions}
We divided the participants into three groups of seven to nine members. The group sizes match real-world work team sizes of external analyst-supported CIM processes \citep{St.Denis2012}. Further, members of geographically distributed teams of up to nine members have been shown to participate more actively and are more committed to and more aware of the team’s goals than in larger teams \citep{Bradner2003}. Our groups were purposefully mixed with participants having complementary skills and expertise so that each group included experts on mapping and data analytics on a similar level. Therefore, the number of participants and the group compositions are a good representation of real-world teams.

\subsection{Scenario design}
The fictional scenario of our experiment was an epidemic outbreak happening simultaneously in three countries. The experiment was inspired by the 2014-2016 Ebola outbreak in Guinea, Liberia, and Sierra Leone. The three country groups had to assess the situation in their respective country by analyzing the data provided during the experiment with the goal to support decisions on where (in which districts) to place treatment centers. The experiment resembled the main challenges of crisis information management, as mentioned in Sections 2.2 and 2.3, by putting participants under time-pressure (urgency), providing incomplete and low-quality data (uncertainty), requiring participants to make high stakes sequential decisions on treatment center placements and having to do so with a shortage of resources.

Before each stage of the experiment, we gave a brief introduction about the scenario and the participants' tasks. Each stage was concluded with a reflection moderated by the researchers.

\subsection{Materials and introduction of representational data bias}
As our participants were experienced practitioners, the data used in the experiment had to resemble reality closely. We used original data from the 2014-2016 Ebola epidemic. The datasets selected for inclusion were on infection rates, infrastructure capacities, demographics, and geography. We adjusted the original data for three reasons. First, some of the participants had been involved in the 2014-2016 Ebola response and should not have a head-start by already being familiar with the data. Second, our experiment required us to introduce a controlled representational bias into the data. Third, the original datasets were too large for the time frame of the experiment. The original data was downloaded from the Humanitarian Data Exchange platform\footnote{\url{https://data.humdata.org/ebola}. Last accessed: October 12, 2021.} and we adjusted it as described in the following.

The \textbf{infection rate} is the key variable in epidemic response. We adjusted the original data so that infection rates were higher and more cases occurred in a shorter time. We retained columns from the original datasets and removed auxiliary columns to avoid information overload in the participants (Table \ref{tab:originaldata}). We included infection data for the first four months of the fictional outbreak (Table \ref{tab:adjusteddata}). Inspired by hidden profile experiments, in our experiment, one district per country was created with substantially more total cases than the other districts in the country. The data of this district was split among group members' datasets (Table \ref{tab:introducedbias}). This implies that only by joining their datasets participants were able to identify the district with the most cases. If the bias remained undetected and untreated, the resulting information products would also become biased.

\begin{table}[]
\caption{Step 1: Retrieving original data from the West-Africa Ebola outbreak. Here truncated to show reported cases of infections. One row is one reported case.}
\label{tab:originaldata}
\resizebox{\textwidth}{!}{%
\begin{tabular}{llllll}
\hline
\textbf{Country} & \textbf{Location} & \textbf{Epi week}                         & \textbf{Case definition} & \textbf{Ebola data source} & \textbf{...} \\ \hline
Liberia          & GRAND BASSA       & 25 to 31 August 2014 (2014-W35)           & Confirmed                & Patient database           & ...          \\
Liberia          & GRAND BASSA       & 08 to 14 September 2014 (2014-W37)        & Probable                 & Patient database           & ...          \\
Liberia          & GRAND BASSA       & 15 to 21 September 2014 (2014-W38)        & Probable                 & Patient database           & ...          \\
Liberia          & GRAND BASSA       & 22 to 28 September 2014 (2014-W39)        & Probable                 & Patient database           & ...          \\
Liberia          & GRAND BASSA       & 13 to 19 October 2014 (2014-W42)          & Confirmed                & Patient database           & ...          \\
Liberia          & GRAND BASSA       & 20 to 26 October 2014 (2014-W43)          & Confirmed                & Patient database           & ...          \\
Liberia          & GRAND BASSA       & 20 to 26 January 2014 (2014-W04)          & Probable                 & Situation report           & ...          \\
Liberia          & GRAND BASSA       & 27 January to 02 February 2014 (2014-W05) & Confirmed                & Situation report           & ...          \\
Liberia          & GRAND BASSA       & 27 January to 02 February 2014 (2014-W05) & Probable                 & Situation report           & ...          \\
Liberia          & GRAND BASSA       & 03 to 09 February 2014 (2014-W06)         & Confirmed                & Situation report           & ...          \\
Liberia          & GRAND BASSA       & 17 to 23 March 2014 (2014-W12)            & Probable                 & Situation report           & ...          \\
...              & ...               & ...                                       & ...                      & ...                        & ...          \\ \hline
\end{tabular}%
}
\end{table}

\begin{table}[]
\centering
\caption{Step 2: Adjusted dataset based on the original data to resemble the infection rate and adapt the data to our fictional country and outbreak.}
\label{tab:adjusteddata}
\resizebox{275pt}{!}{%
\begin{tabular}{lllll}
\hline
\textbf{Country} & \textbf{District} & \textbf{Month} & \textbf{Case definition} & \textbf{Ebola data source} \\ \hline
Noruwi           & Aameri            & 1              & Confirmed                & Situation report           \\
Noruwi           & Aameri            & 1              & Probable                 & Situation report           \\
Noruwi           & Aameri            & 1              & Probable                 & Patient database           \\
Noruwi           & Aameri            & 2              & Probable                 & Situation report           \\
...              & ...               & ...            & ...                      & ...                        \\
Noruwi           & Aameri            & 3              & Confirmed                & Patient database           \\
Noruwi           & Aameri            & 4              & Probable                 & Patient database           \\
Noruwi           & Aameri            & 4              & Probable                 & Situation report           \\
Noruwi           & Aameri            & 4              & Probable                 & Situation report           \\ \hline
\end{tabular}%
}
\end{table}

\begin{table}[]
\caption{Step 3: Introduction of representational bias. We created biased versions of the adjusted datasets from step 2. The biased versions were distributed among participants. The bias is here introduced in the district of \textit{Niprusxem}. The district has the most cases in the unbiased dataset, but the least cases in the biased datasets. One group member only receives data for month 1 (displayed). Each other group member also only receives data for one month (not displayed). Only by joining the datasets, the unbiased case numbers could be received.}
\label{tab:introducedbias}
\resizebox{\textwidth}{!}{%
\begin{tabular}{lllllllllll}
\hline
\textbf{}          & \multicolumn{5}{c}{\textbf{Unbiased}}                                                                                                                                      & \multicolumn{5}{c}{\textbf{Biased}}                                                                                                                                        \\ \hline
\textbf{Districts} & \multicolumn{1}{c}{\textbf{M1}} & \multicolumn{1}{c}{\textbf{M2}} & \multicolumn{1}{c}{\textbf{M3}} & \multicolumn{1}{c}{\textbf{M4}} & \multicolumn{1}{c}{\textbf{Total}} & \multicolumn{1}{c}{\textbf{M1}} & \multicolumn{1}{c}{\textbf{M2}} & \multicolumn{1}{c}{\textbf{M3}} & \multicolumn{1}{c}{\textbf{M4}} & \multicolumn{1}{c}{\textbf{Total}} \\ \hline
Aameri             & 4                               & 12                              & 44                              & 140                             & 200                                & 4                               & 12                              & 44                              & 140                             & 200                                \\
Baldives Saintman  & 3                               & 21                              & 27                              & 147                             & 198                                & 3                               & 21                              & 27                              & 147                             & 198                                \\
Bana Cadi          & 1                               & 2                               & 24                              & 54                              & 81                                 & 1                               & 2                               & 24                              & 54                              & 81                                 \\
Grethernquetokong  & 1                               & 8                               & 12                              & 52                              & 73                                 & 1                               & 8                               & 12                              & 52                              & 73                                 \\
Janmantho          & 1                               & 6                               & 19                              & 39                              & 65                                 & 1                               & 6                               & 19                              & 39                              & 65                                 \\
Lemau              & 4                               & 4                               & 92                              & 140                             & 240                                & 4                               & 4                               & 92                              & 140                             & 240                                \\
Mau Cari           & 1                               & 4                               & 20                              & 49                              & 74                                 & 1                               & 4                               & 20                              & 49                              & 74                                 \\
Menia              & 1                               & 1                               & 20                              & 32                              & 54                                 & 1                               & 1                               & 20                              & 32                              & 54                                 \\ \hline
\textit{Niprusxem} & \textit{5}                      & \textit{20}                     & \textit{125}                    & \textit{160}                    & \textit{310}                       & \textit{5}                      & \textit{0}                      & \textit{0}                      & \textit{0}                      & \textit{5}                         \\ \hline
Samac Iali         & 1                               & 3                               & 17                              & 62                              & 83                                 & 1                               & 3                               & 17                              & 62                              & 83                                 \\
Southdos Dinia     & 3                               & 12                              & 66                              & 129                             & 210                                & 3                               & 12                              & 66                              & 129                             & 210                                \\
Thesey             & 1                               & 3                               & 24                              & 37                              & 65                                 & 1                               & 3                               & 24                              & 37                              & 65                                 \\
Usda Nilia         & 1                               & 4                               & 14                              & 29                              & 48                                 & 1                               & 4                               & 14                              & 29                              & 48                                 \\
Walof              & 1                               & 2                               & 12                              & 42                              & 57                                 & 1                               & 2                               & 12                              & 42                              & 57                                 \\
Total              & 28                              & 102                             & 516                             & 1112                            & 1758                               & 28                              & 82                              & 391                             & 952                             & 1453                               \\ \hline
\end{tabular}%
}
\end{table}

\textbf{Infrastructure and capacity data}:
During the 2014-2016 Ebola outbreak, mapping healthcare facilities and their capacities became a crucial task for crisis information management. However, up to 60~\% of values in the original data on health infrastructure and capacities were missing, highlighting once more the high uncertainty analysts are confronted with. In addition, values had unclear and ambiguous meanings, making interpretation difficult. We adjusted the original datasets to include a reduced number of key variables. In the original datasets, detailed capacity data, i.e., numbers of beds per treatment center, was incomplete for 58~\% of entries. We mimicked this representational bias in our adjusted datasets. Only one participant per group received capacity data on the number of beds per facility. The other group members received the same dataset but with an empty column for capacities.

\textbf{Demographic and geographic data}:
Demographic data are part of the common operational datasets in crisis response \citep{VandeWalle2010}. They are used to understand the overall population distribution in terms of age, gender, and geographic location. By providing a sense of population density and bordering regions, they become very important in predicting trends in epidemic outbreaks. We collected the original data, replaced country and district names with randomly generated names, and slightly adjusted the demographic numbers. We further included randomly generated maps corresponding to the three randomly generated countries and districts. The maps were distributed to the participants in digital and printout versions.

\textbf{Data volume}: Data volume differed slightly between the groups, with no large differences that could have significantly eased or complicated one group's data review and analysis process (Table \ref{tab:datadimensions}).

\begin{table}[]
\centering
\caption{Dimensions of datasets handed to groups. Dimensions given in rows x columns.}
\label{tab:datadimensions}
\resizebox{150pt}{!}{%
\begin{tabular}{lll}
\hline
\textbf{Group} & \textbf{Dataset} & \textbf{Dimensions} \\ \hline
Noruwi         & Infection cases  & 1759x4                               \\
Noruwi         & Demographics     & 15x22                                \\
Noruwi         & Capacity         & 58x19                                \\
Reloupe        & Infection cases  & 1724x4                               \\
Reloupe        & Demographics     & 14x5                                 \\
Reloupe        & Capacity         & 64x19                                \\
Republic       & Infection cases  & 3142x4                               \\
Republic       & Demographics     & 36x22                                \\
Republic       & Capacity         & 87x19                                \\ \hline
\end{tabular}%
}
\end{table}

\textbf{Participants' access to the data}:
We created Google accounts for each participant, and the created datasets were uploaded into the Google Drive folders of each participant. This allowed us to distribute the created datasets to the members of each group while making sure the introduced bias was identifiable. A print-out sheet with login information for the Google folder was created for each participant. Each participant received a laptop to access the files. The laptops had MS Office pre-installed for the information management work on the data. Further tools also used by our participants in their professional work, including RStudio Online and Google Spreadsheets, were also available.

\subsection{Experimental setup and procedure}
To address the first two research questions (\textit{Is surging external analysis capacity effective in identifying and mitigating data bias?} and \textit{How do external analysts and decision-makers jointly handle data bias in the decision process?}), we set up the first two stages of the experiment. To address research question three (\textit{Does confirmation bias create path dependencies whereby biased assumptions persist in sequential decisions?}), we conducted an online survey with the same participants.

\subsubsection{Experiment stage 1}
Stage 1 was conducted only with the group of external analysts. They were divided into the three groups we had defined in the planning of the experiment (Table \ref{tab:participants}). Each group was responsible for the information management for one country affected by the fictional outbreak.

Participants were told their group's objective was to review the available data and develop information products that could be used in stage 2 of the experiment for the prioritisation of districts that needed most urgent assistance. As all participants are used to preparing information products for crises, they were free to decide which information products to create (e.g. maps, tables, graphs, etc.). Participants were briefed they could use the MS Office Suite installed on the laptops provided to them, or any other online tools they would use in their professional work. Because of participants' experience, the importance of developing accurate information was clear to them. This includes the checking of data issues, gaps and comparing information quality among group members. We gave them no indication that they could expect the data they received to be perfect, accurate and unbiased. Rather, we briefed them that the experiment should be seen as a simulation of a real case, with challenges that can be expected from real epidemic crises. Participants were briefed they had 2.5 hours for their task.

After the introduction, the three groups formed in three rooms, equipped with laptops and information sheets that contained user-login information for each participant to access the available data. The groups were asked to present the developed information products and suggestions for response decisions at the end of experiment stage 1.

\subsubsection{Experiment stage 2}
In stage 2, decision-makers joined each of the three groups. Participants were briefed they had to make resource allocation decisions by placing treatment centers in priority districts of their respective countries. External analysts had to brief the decision-makers on the outbreak situation, priority issues, and districts using the information products developed by them in stage 1. Each group received a limited amount of treatment centers (in the form of small building blocks) that could be placed in districts of the fictional countries on printout maps. Participants were told that each treatment center, i.e., building block, had a fixed capacity of ten beds. We implemented resource constraints by limiting the number of available treatment centers and beds. Thus, not all districts could be fully equipped to respond to the rising infections and prioritization decisions had to be made. Participants were briefed that all decisions had to be made within 60 minutes.

After the introduction, the three groups formed in three rooms, equipped with laptops and the information products developed in stage 1. The groups were asked to present their final decisions at the end of the experiment.

\subsubsection{Experiment stage 3}
To address the third research question after stage 2 was completed, all participants were asked to fill out an online survey on site. The research objective was to assess whether confirmation bias would lead to path-dependencies toward decisions that were made based on biased information. A significant confirmation bias result would mean that participants preferred to seek information that confirmed their previously formed assumptions, even when they were influenced by biased datasets. 

The survey referred to participants' previous decision from stage 2, where they selected a priority district to which most treatment centers were allocated. In stage 3, participants were briefed that new information was available after they had made prioritization and allocation decisions. Their task was to select from a list of datasets those ones that they found most important to support further information management and decision-making. The survey item and confirmation bias measure is described in section \ref{lab:cbdatacollectionanalysis}.

\subsection{Data collection and analysis}
\subsubsection{Experiment stage 1 and 2}
In stages 1 and 2, one observer per group took notes of the information management processes, communication, and interaction within the groups. Photos were taken to document intermediate results and processes, for example of post-its on the printout maps. After the session, the group members' files of the information products created on the laptops were saved and analyzed by the researchers. 

We conducted structured observations of the first two stages of the experiment that included the use of protocol sheets with guiding questions. Data collection through researcher observation is highly suitable in interactive experimental settings with dynamic group discussions. The goal was to capture verbal data, i.e., what is discussed, how by whom and when, as well as interactions among group members \citep{steffen2019einfuhrung}. Since an observer must select which person and interaction is the object of observation (selection problem), a result bias can occur \citep{steffen2019einfuhrung}. We addressed this potential issue by briefing observers beforehand on the observation protocol and guiding questions. Thus, before beginning an observation, researchers numbered participants in a common format to protocol activities in a standardized way, quickly and effectively. The protocol guideline included example observation items and was divided into three different sections: (1) description of workshop site, (2) communication and interaction description, (3) general impressions. The complete observation protocol is provided in the Appendix. The collected data was evaluated through qualitative content analysis \citep{doring2016forschungsmethoden}. The main activity was to summarize the collected observational data and reveal content related to our research questions. We further evaluated the information products developed by the participants in addition to conducting the qualitative document analysis. We proceeded in three steps:

\begin{enumerate}
\item \textbf{Paraphrasing:} To reduce the volume and complexity of the observational data and of the created information products, the first step was to identify passages that carry content relating to our research questions and delete passages that did not. In this process, the different data forms (text passages of the sheets and information products, e.g. maps) were analyzed separately.
\item \textbf{Coding:} In the second step, all paraphrases representing the main content were summarized in a single document. The separate paraphrases were coded and structured to answer our research questions and find explanations for these answers. We conducted two coding iterations to develop a set of coded categories of the observed discussions and activities.
\item \textbf{Analyzing:} In the final step, we analyzed the structured content with regard to our research questions. Through this content analysis, we were able to systematically evaluate and analyze all observation sheets and information products and present key results.
\end{enumerate}

The first author coded the data in the first iteration. The resulting codes and corresponding observational notes were discussed with the second author. Adjustments were made to some of the coded categories, followed by the second iteration of coding by the first author. After review by the whole author team, the final categories of codes were agreed on. Table \ref{tab:observationnotes} presents example observation notes and coded categories.

\begin{table}[H]
\caption{Example observation notes taken during the experiments and respective coded categories.}
\label{tab:observationnotes}
\resizebox{\textwidth}{!}{%
\begin{tabular}{lc}
\hline
\multicolumn{1}{c}{Example   observation notes}                                                                                      & Coded category                         \\ \hline
Express need for information: transportation network                                                                                                  & Requirements for additional data       \\
\begin{tabular}[c]{@{}l@{}}Discussing data gaps: more background data on the\\  country, transmission data, spread on daily basis needed\end{tabular} & Requirements for additional data             \\
Should we merge our data?                                                                                                                             & Debias behavior                                              \\
\begin{tabular}[c]{@{}l@{}}Questioning why they have different datasets. \\ Trying to understand the cause of the data bias\end{tabular}              & Debias behavior                                            \\
\begin{tabular}[c]{@{}l@{}}One person uploaded their files into a shared folder, \\ all others used the data from there\end{tabular}                  & Data sharing                           \\
Receiving data from other groups                                                                                                                      & Data sharing                           \\
\begin{tabular}[c]{@{}l@{}}Deliberation of format of final information product\\ for decision support\end{tabular}                                    & Discussion on decision recommendations  \\
\begin{tabular}[c]{@{}l@{}}Information product proposal: curve by day, what \\ is happening, did people die or not\end{tabular}                       & Discussion on decision recommendations  \\
Using familiar tool to create digital, layered map                                                                                                    & Data work                              \\
Creation of (biased) aggregates for numbers of cases                                                                                                  & Data work                                                                            \\

Not sure what the most important dataset is                                                                                                           & Interpretation of data                 \\
Need to know: where is the death rate the highest?                                                                                                    & Interpretation of data                 \\
the data is not   very clean; possibly underreporting                                                                                                 & Communicating data limitations         \\
we had different datasets between group   members                                                                                                     & Communicating data limitations         \\

Decision-makers   studying the developed map                                                                                                           & Interpretation of situation            \\
Discussion   possible causes for the outbreak                                                                                                         & Interpretation of situation            \\
Need to   make a decision; what do we have and what is missing                                                                                        & Allocation strategy                    \\
where   NOT to put centres?                                                                                                                           & Allocation strategy                    \\
Communication   of available recources/capacities                                                                                                     & Discussing capacities                  \\
Clarification   of center capacities                                                                                                                  & Discussing capacities                  \\ \hline
\end{tabular}%
}
\end{table}

\subsubsection{Experiment stage 3}
\label{lab:cbdatacollectionanalysis}
In stage 3, participants were asked to complete the online survey on site. The survey was implemented in a Google Form and distributed to each participant. The survey prompted the participants with the following text: \textit{"Below are the summaries of 10 new datasets that are available. You can request the full version of those datasets but you only have limited time and resources to evaluate them all in detail. Select as many datasets as you want. District X is the district you have identified in the last session as the most critical district."} 

In stage 2, participants had to allocate treatment centers to the districts with the highest priority (referred to as \textit{"District X"} in the survey). In the survey, ten summaries of ten fictional datasets were given in one-sentence statements. Five dataset summaries supported that District X was indeed a priority district, whereas the other five dataset summaries opposed this. An example of a summary of a supporting dataset is \textit{"Dataset 9: District X has a high amount of health care workers infected."} An example of a summary of an opposing dataset is \textit{"Dataset 10: District X has a low amount of heath care workers infected."} 

Participants did not receive any data to review besides those summaries, and after the survey was completed, they did not receive the datasets they selected, as it was not necessary to measure confirmation bias \citep{Jonas2001, Fischer2011}. The complete confirmation bias measure can be found in the Appendix.

The response data from the survey was imported into SPSS for statistical analysis. Following the measures of confirmation bias in previous studies, we first counted the selected supporting and opposing datasets per participant. Then, we used a paired samples test to identify whether the mean counts of selected confirming and opposing datasets were significantly different.

\section{Results}
\label{sec:results}
In the following, we present the results for our three research questions.

\subsection{Impact of external analysis capacity on data biases}\label{sec:res_stage_1}

In the first stage of the experiment, all three groups of external analysts identified differences between group members’ datasets and discovered that the data providing the numbers of infections were biased.

\textbf{Example observation: }\textit{EA8 is looking up the data for Niprusxem. He says he only has month 2 for this and that this is strange. Asks to see EA12’s data. EA9 says she only has month 3. EA12 has month 4.
EA9: We have different datasets!}

However, the bias within the capacity data remained undetected in all three groups (see Table \ref{tab:identifieddatabiases}). This led to the development of information products that were overly focused on the outbreak situation and overlooked existing capacities.

\begin{table}[H]
\centering
\caption{Overview of identified bias per group}
\label{tab:identifieddatabiases}
\resizebox{250pt}{!}{%
\begin{tabular}{lcc}
\hline
\textbf{Group} & \multicolumn{1}{l}{\textbf{Bias in infection data}} & \multicolumn{1}{l}{\textbf{Bias in capacity data}} \\ \hline
Noruwi         & Identified                                                       & Not identified                                                       \\
Reloupe        & Identified                                                       & Not identified                                                       \\
Republic       & Identified                                                       & Not identified                                                       \\ \hline
\end{tabular}%
}
\end{table}

\begin{figure}[h]
\caption{Experiment stage 1 results of the coding and analysis process. The figure shows the share (in percentage over time) of the coded categories within the overall activities of the groups. Debiasing efforts were not sufficiently followed up upon and, towards the end of the experiment, largely replaced by discussions on decision-making recommendations.} 
\vspace*{+2mm}
\label{fig:experiment1results}
\centering
\includegraphics[width=\textwidth]{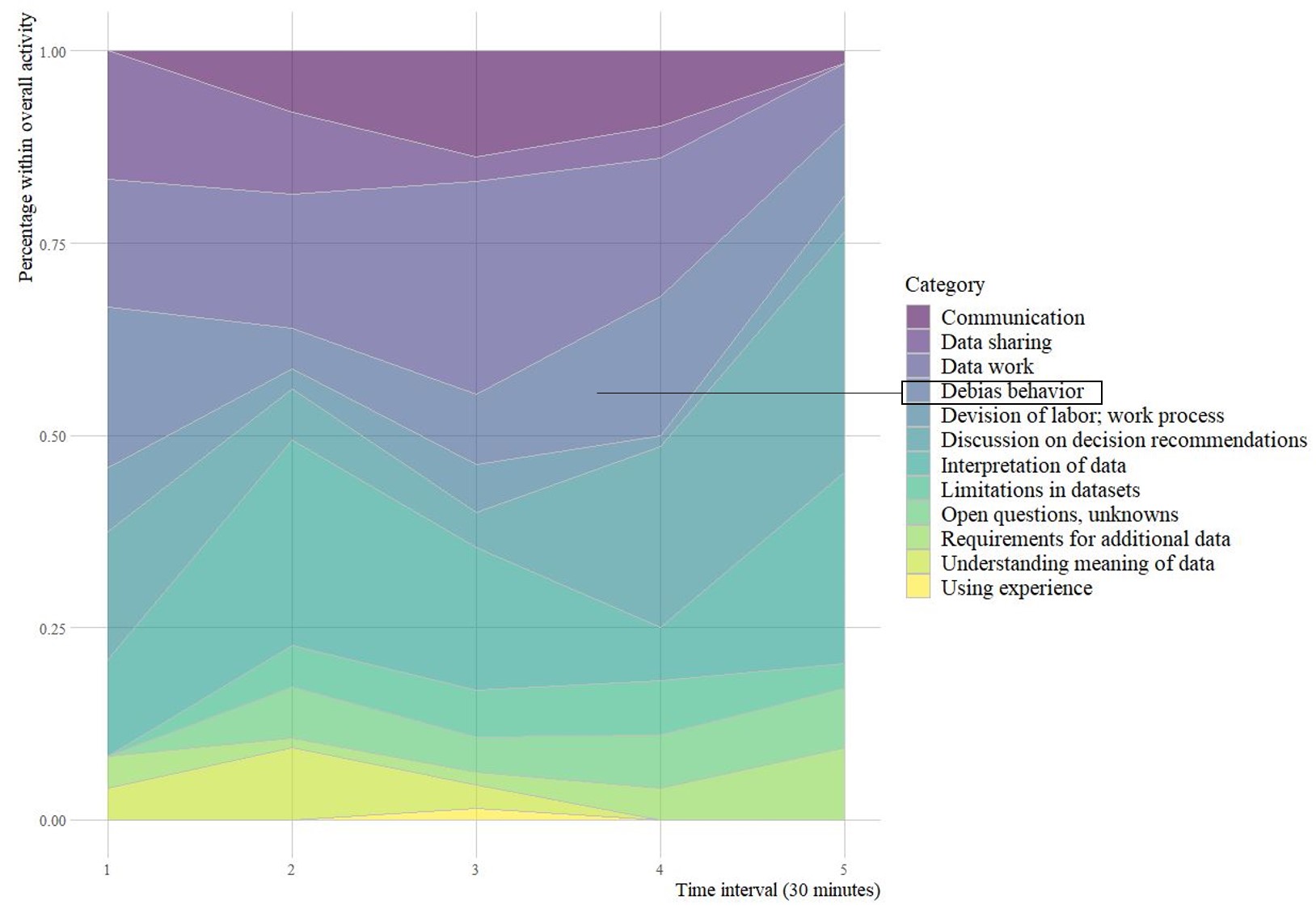}
\end{figure}

Figure \ref{fig:experiment1results} shows the results of the coding and categorization process of our qualitative content analysis. The figure provides a summary of the sensemaking process within the groups. It shows the share of each coded category (in percent) within the overall activities of the groups during five time intervals of 30 minutes each. 

In the initial phase, participants rushed into downloading the datasets stored in their individual Google accounts and started the data analysis by importing the data into their preferred information systems (e.g., Excel, RStudio). Participants  familiarized themselves with their own data and identified differences in the data of their group members. Figure \ref{fig:experiment1results} shows the share of \textit{data work} remained constant during the first two time intervals (i.e. first 60 minutes). It became the dominant category during the third interval and then lost importance by making room for an increased focus on \textit{decision-making recommendations}. Figure \ref{fig:experiment1results} also shows the groups started with attempts to integrate datasets as \textit{debiasing behavior} in the first interval. 

\textbf{Example observation: }\textit{EA10 suggests to the group to upload the data into Google Drive so he can easily merge them.}

These attempts were, however, not efficiently followed-up upon, and the share of \textit{debiasing behavior} was reduced in the second time interval.

After an initial familiarization with the data, a collective sensemaking process started to emerge, characterized by intensive socializing, working, and experimenting with the data. The groups discussed how to define priority districts and what should be the key variables. This led to \textit{debiasing behavior} gaining significance slightly and reaching its peak at the second last interval when groups recognized that datasets remained biased. The sensemaking process did not lead to due attention to biases. When differences between the group members’ datasets were recognized, measures taken by the groups were insufficient to debias the data. One reaction was that one group member would upload their biased dataset into a shared folder, and the other group members would from then on use this data folder as the single-point-of-truth. From that time on, all group members accessed the same biased data. This behavior might be explained by groupthink, as the individual members of the groups strived to establish harmonic relationships, characterized by conformity and the minimization of conflict rather than openly articulating the disconfirming information they held.

Participants struggled with the non-availability of data they wished to have and perceived the data quality of some datasets to be too low to build accurate situational awareness and determine priorities. With the end of the experiment stage approaching and time pressure increasing, groups tasked individual members with creating information products, i.e., maps, graphs, and tables. 

\textbf{Example observation: }\textit{EA11: Data quality is questionable, it is not meaningful to go into data analysis in the last 20 minutes, must be quick... I need to think of the report, we should still name projects or tasks that our organizations would work on.}

At this point, it became increasingly difficult for the groups to mitigate any data biases because individuals would turn their own data into information for decision support, and no critical data assessments were done. Figure \ref{fig:experiment1results} shows \textit{Interpretation of data} and \textit{decision-making recommendations} dominated the last time interval and \textit{debiasing behavior} was again neglected.

\begin{figure}[tb]
\caption{Example information product resulting from stage 1. Country map shows the numbers of cases per district (colored by the participants in red, yellow, and blue). The green box (added by us) shows that the unbiased numbers of cases for the most affected district were much higher than those reported in the information product developed by the participants.}
\label{fig:exampleinfoproduct}
\centering
\includegraphics[width=\textwidth]{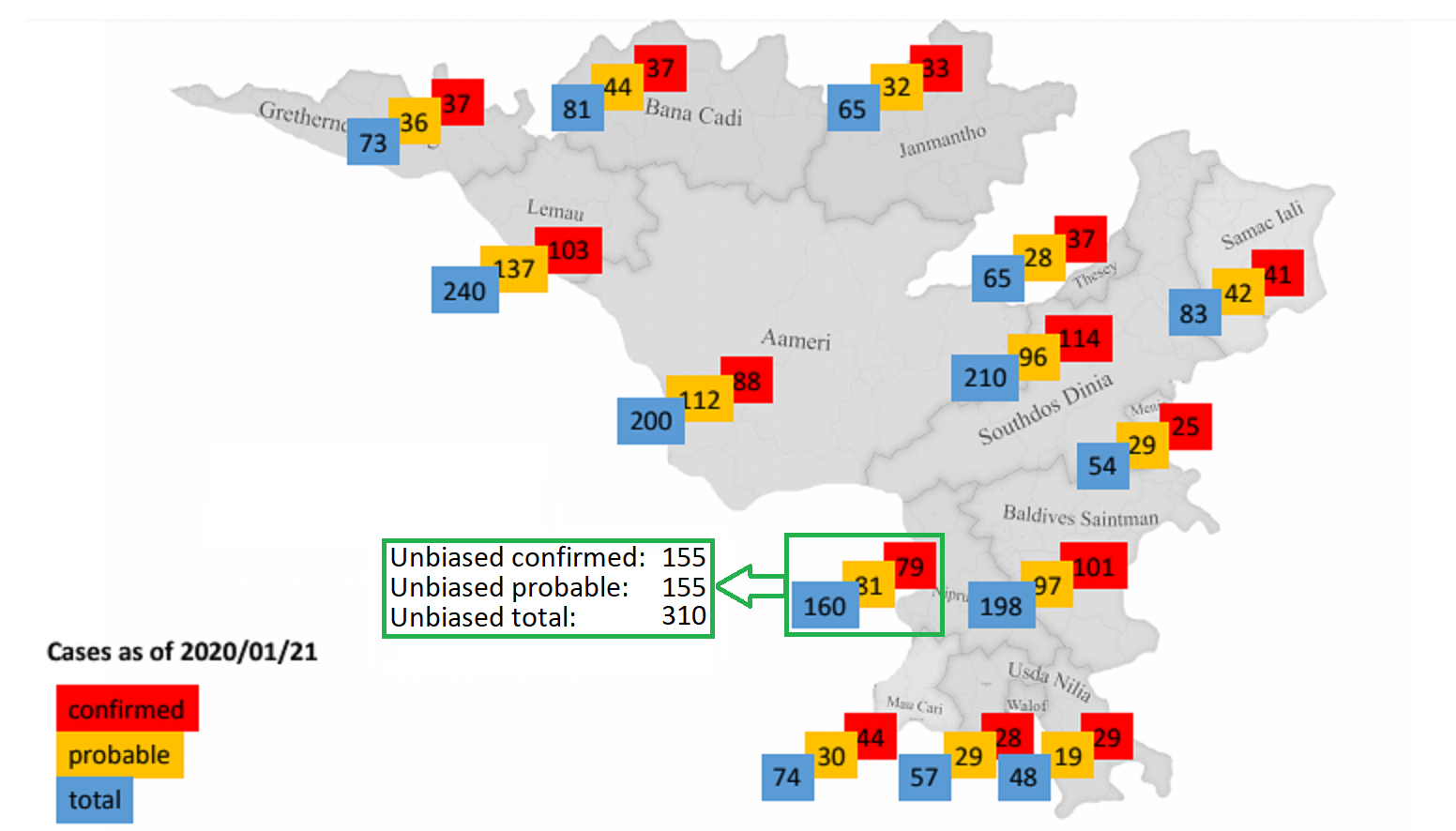}
\end{figure}

Even though all groups identified the bias within the infection data, the groups failed to successfully debias the data. Successful debiasing would have required that members of each group merge their datasets for infection rates and infrastructure capacity. However, even though the bias was recognized, each group relied on the data of only one of its members in the design of information products. Remarkably, one group identified early during the experiment that its members had received biased data and shared their finding with the other groups, but still all groups presented results based on biased data at the end of the experiment. 

The resulting information products of each group showed numbers of infections in the most affected districts that were lower than the complete and unbiased information they could have acquired by merging their datasets. Figure \ref{fig:exampleinfoproduct} shows one example of a developed information product. It depicts that the district with the most cases in the unbiased dataset was presented with biased numbers based on only one of the participants’ datasets.

Overall, an explanation for the unsuccessful debiasing is the strong perception of time pressure and the experienced urgency by participants to deliver an information product in time that is presentable and actionable for decision-makers. Even though the additional analyst capacity is meant to alleviate the time pressure, they are subject to the same biases of exploiting, rather than exploring data  \citep{Comes2020}. Analysts were not able to develop unbiased information products for decision support, since the data was accepted with its flaws, and information products needed to be developed anyway based on the low-quality data.

\subsection{Data bias in the decision processes} 

In experiment stage 2, all three groups relied on the biased datasets and resulting biased information products from stage 1 in their discussions on treatment center placement decisions. External analysts briefed decision-makers using the biased numbers of infections. 

\textbf{Example observation: } \textit{They decide to place treatment centers based on the case numbers, and also want to place them along the border. EA12 shows the map of the confirmed cases to the DMs.}

As described Section~\ref{sec:res_stage_1}, no group was able to identify the data bias on existing bed capacities during information product development (see Table \ref{tab:identifieddatabiases}). Consequently, no detailed capacity data was communicated to decision-makers, and allocation decisions were made in the absence of detailed data on existing capacities. If the capacity data bias had been discovered, it potentially could have facilitated the groups’ allocation decisions. 

Decision-makers took the role of \textit{advocatus diaboli} by critically questioning the underlying data of the developed information products. In their role as decision-makers, they pressured external analysts on the data gaps and data quality issues very early in the experiment. 

\textbf{Example observation: } \textit{DM3: why are some areas empty? EA5: the data is not very clean; possibly underreporting. DM1: is the data trustworthy? EA5: we had different datasets between group members.}

Analysts briefed decision-makers on data limitations. This led to the joint understanding that the available data was unreliable to some degree. However, when data limitations were mentioned, decision-makers did not pressure enough. When analysts explained data gaps, other group members, who had access to that missing data, would not step in to clarify. Decision-makers would not press the group sufficiently to mitigate the data bias. Instead, they would pressure to make prioritization decisions for treatment center allocation.

\textbf{Example observation: } \textit{DM5: Based on my experience, you have to make decisions on very little data. Indecision kills.}

\begin{figure}[h]
\caption{Experiment stage 2 results of the coding and categorization process. The graph shows the share (in percentage over time) of the coded categories within the overall activities of the groups. Initial discussions on data limitations were not sufficiently followed-up upon and discussions on allocation strategy dominated the group discussions from the second interval onward.}
\vspace*{+2mm}
\label{fig:experiment2results}
\centering
\includegraphics[width=\textwidth]{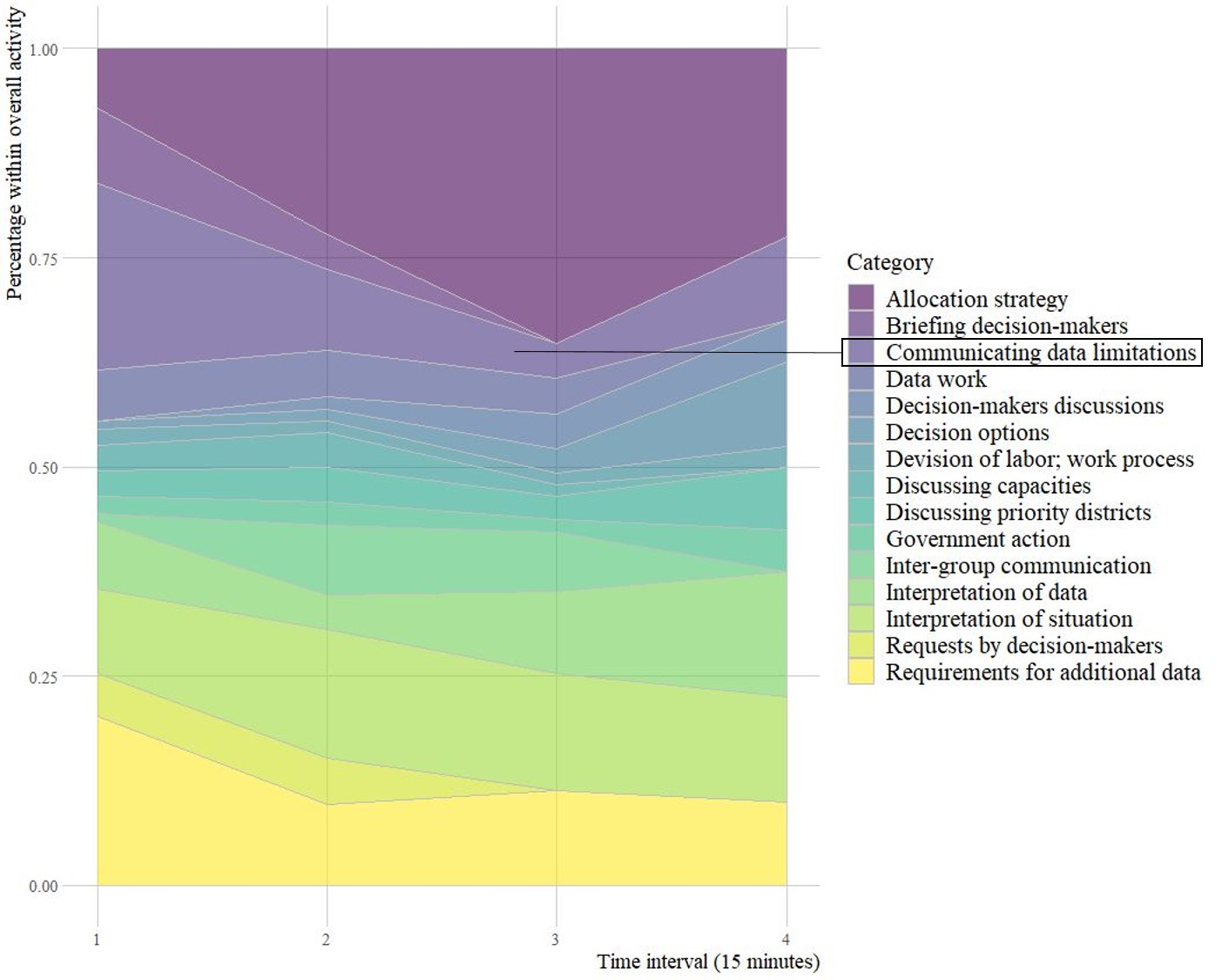}
\end{figure}

Figure \ref{fig:experiment2results} shows the results of the coding and categorization process of our qualitative content analysis of experiment stage 2. It shows the share of the coded categories (in percent) within the overall activities of the groups during four time intervals which are 15 minutes each. Deliberations on \textit{allocation strategies} dominated discussions from the second interval onward till the end of the experiment. It reached its peak during the second last interval, where 35~\% of discussions were on \textit{allocation strategies}.

Groups showed stronger debiasing behavior at the beginning of the session, where \textit{data limitations} were communicated and discussed. However, this focus was reduced over time, only increasing slightly in the last time interval. This pattern of debias neglect was already observed in stage 1.

\textit{Requirements for additional data} mainly were articulated in the beginning and were then constant throughout the later intervals even though it was communicated to the participants that there would be no additional data provided during the experiment. This behavior shows a heavy dependency on \textit{more} data and the conviction that more data will help the decision process, even if the quality of the future data is unknown and can be questioned if the currently available data is already of low quality.

\textit{Interpretation of the situation} out-weighted the \textit{interpretation of the data} throughout all intervals, showing the influence of the decision-makers who relied more on their previous experience to assess the situation than basing their assumptions on the available data that was known to have limitations.

Overall, the joint information management and decision-making process between analysts and decision-makers did not result in sufficient debiasing, and allocation decisions were made based on biased information.

\subsection{Persistence of bias in sequential decisions}

In the final phase, participants were asked to select additional information that supported or conflicted with their allocation decisions. Our analysis of the survey responses shows that the mean count of selected supporting datasets was higher (M = 2.94, SD = 1.56) than the mean count of selected opposing datasets (M = 1.82, SD = 1.88), indicating that participants selected more supporting than opposing datasets (see Table \ref{tab:descriptivestats}). Wilcoxon signed-rank test was used to test if the discrepancy between means was statistically significant. The result reveals significant confirmation bias in the participants’ selection of additional datasets (n = 17, z = -2.537, p = .011). We, therefore, find that our participants showed significant confirmation bias and that the bias drives their information selection decisions. 

This is particularly concerning as the participants' preliminary decisions were flawed and based on biased information. In stage 3, participants tried to substantiate further their previously biased decisions instead of using the opportunity to counter-check their assumptions. Confirmation bias reinforced their biased assumptions and strengthened their reliance on potentially further biased data. 

A significant confirmation bias at this stage is in line with our observations in the earlier stages of the experiment, where participants followed on exploitative and satisficing strategy given the time pressure, rather than an exploratory strategy. Although much of the literature on crisis and disaster management suggests an adaptive approach to manage the uncertainties that typically persist at the onset of a crisis \citep{Comes2020, quarantelli1988disaster}, we found that over time the initial mental models and decisions became deeply ingrained and persistent. As such, it became increasingly difficult for participants to implement a debiasing strategy that allowed them to correct their decision because the initial data biases were never effectively discussed and mitigated, even though new information became available that could have facilitated corrections. Even though they knew that their information had been incomplete and possibly flawed, the participants' debiasing behaviour was diminished, and they were overconfident in their decisions. If participants would have laid more focus on discussions on data limitations, they might have been more mindful and showed a more balanced or even disconfirming information selection behavior to correct previously flawed decisions.

\begin{table}[H]
\centering
\caption{Descriptive statistics of Wilcoxon-Singed-Rank Test.}
\label{tab:descriptivestats}
\resizebox{300pt}{!}{%
\begin{tabular}{llllll}
\hline
           & N  & Mean & SD   & \begin{tabular}[c]{@{}l@{}}Count of selected\\ datasets (min)\end{tabular} & \begin{tabular}[c]{@{}l@{}}Count of selected\\ datasets (max)\end{tabular} \\ \hline
Opposing   & 17 & 1.82 & 1.88 & 0                                                                          & 5                                                                          \\
Supporting & 17 & 2.94 & 1.56 & 0                                                                          & 5                                                                          \\ \hline
\end{tabular}%
}
\end{table}

\section{Discussion}
\label{sec:resultslimitations}

\subsection{Contribution to literature}

Our experimental evidence adds to the theoretical understanding of the role of biases and debiasing strategies in crisis information management \citep{Mirbabaie2020, Ogie2018, comes2016cognitive}. Our experiments show that a reason for the lack of debiasing efforts is the urgent context of crisis information management and strong group cohesion lead to a neglect of critical data assessments within the initial exploratory step of the analysts. Debiasing behavior is particularly strong during the onset of workgroup collaborations. However, these debiasing efforts are increasingly neglected as time pressure builds and mental models are formed. This implies that rather than using additional capacity to broadly scan the available information, the process follows a satisficing strategy, where by one data set is 'good enough' to develop information products quickly that are directly actionable and support decision-making. Because biases remain untreated, information products and decisions become affected by them.

Even though conventionally there is hope that additional data analysts will mitigate the impact of data bias, our findings show that even though biases are detected, they are not mitigated. \cite{Hughes2015} emphasized the expertise of external analysts with specialized software. We find that the preference to start data analysis quickly in participants' preferred tools moves the focus away from debiasing efforts. The law-of-the-instrument was clearly present in our groups, especially in the initial phase of the experiment. This indicates that our participants had strong preferences for their preferred information systems. In an effort to understand their own data, participants approached  data analysis with tools they were familiar with and knew best. Datasets from other group members, and their potential differences, were not receiving due attention.

Our findings show the interplay of data and cognitive bias in crisis response. We find that confirmation bias can exacerbate the reliance on biased assumptions and that data biases and cognitive biases can reinforce each other, leading to amplified bias effects. As proposed by \cite{comes2016cognitive}, and experimentally confirmed in our study, crisis information managers and decision-makers are prone to significant confirmation bias. Our participants significantly more often selected new information that confirmed their previous assumption about priority districts, which was influenced by biased data. This holds true even considering the broad level of experience of our participants, and although they did know the initial data was biased. We therefore show that awareness of bias does not automatically lead to bias mitigation. The urgent, uncertain, and resource-constraint contexts of crisis response have led to calls for adaptive management \citep{Merl2009, Janssen2020, Charles2010, Anson2017, schiffling2020implications, turoff2004design}. Our findings indicate that such adaptive approaches can fail due to the interplay of data and cognitive bias.

\subsection{Mindful debiasing and future research}

Future CIM theory needs to further explain the interplay of data bias and cognitive bias, looking into reinforcing and mitigating mechanisms. Crisis situations are known to cause stress in responders, and this stress is known to increase the susceptibility to cognitive biases such as confirmation bias. Especially in data-critical environments like CIM, where responders have to handle various information systems, techno-stress can further increase stress and susceptibility to biases. Mindfulness has been found to alleviate some of this stress \citep{Ioannou2017} and therefore is a promising strategy to reduce the susceptibility to cognitive bias in CIM. Mindfulness means being more aware of the context and content of the information one is engaging with \citep{Langer1992}. When crisis information managers are mindful about the context and content of the information they are engaging with, falling into the trap of ever-confirming information-seeking behavior becomes less likely. In a mindful state, information managers would be more open to new and different information, and able to develop new categories for information that is received. In contrast, in a less mindful state, people rely on previously constructed categories and neglect and ignore the potential novelty and difference within newly received information. Being mindful means to increase one's metacognition, i.e., being aware and have a focus on one's own thought processes \citep{Croskerry2013}. Boosted metacognition might be effective in mitigating confirmation bias \citep{Rollwage2021}. Future research should investigate the effectiveness of such debiasing efforts empirically.   

Like \cite{Ogie2018}, we argue that data created in crises, especially from the affected population, can be subject to a multitude of biases, which have to be taken into account if systems and algorithms are designed that are supposed to turn those data into objective, neutral decision recommendations. In a similar vein as \cite{Weidinger2018}, who called for more research on users' perception of novel information systems and technologies for crisis response, we argue, information management literature needs to account for data biases that systematically over- or under-represent issues, social groups, or geographic areas in the form of representational biases. If information management does not account for biases, resulting information products can become flawed and negatively influence decision-making. 

Previous research proposed new forms of information systems, models, and algorithms to support resource allocation decisions in crises \citep{Avvenuti2018, Kamyabniya2018, schemmer2021conceptualizing}. We argue that such systems need to consider the abilities and limitations of information managers and decision-makers to identify and mitigate biases in the usage of such systems. This includes data biases as well as cognitive biases. We emphasize previously proposed debiasing efforts, e.g., nudging \citep{Mirbabaie2020}, that can be implemented into information systems for crisis response with the objective to mitigate cognitive biases.

Previous research provided examples on  effective debias interventions. Interventions can range from fast and frugal options to intensive training sessions \citep{Sellier2019}. Information managers and decision-makers can be trained to counter-check their assumptions by actively seeking disconfirming information and considering the opposite of their preliminary hypothesis \citep{Satya-Murti2015, Liden2019}. Future research needs to test the effectiveness of such interventions in crisis settings. 

We reiterate calls for sensemaking support in crisis response \citep{Muhren2010, Comes2020}. We add to that with our finding that decision-makers can act as advocatus diaboli to their external analyst partners. By trying to make sense of the unfolding situation and posing confrontational questions to external analysts regarding the quality and shortcomings of the data that underpinned developed information products, decision-makers uncovered important data gaps quickly. However, these also have to be effectively followed-up upon to lead to lead to successful debiasing.

\subsection{Implications for practice}

It can be observed that the response organizations are building up stronger internal crisis information management structures. Where once there were large skill gaps in data analysis and mapping, digital response concepts are now being observed within established organizations \citep{Fiedrich2021}. External analysts are being integrated into permanent structures. 

However, our findings suggest that crisis information management needs to invest in detecting, and most importantly, mitigating biases. Even if complete debiasing is not feasible, we give some concrete implications of our findings on crisis information management practice.

First, bias-awareness trainings can highlight the potential influence of biases in information management and decision-making, and provide guidelines for debiasing. We found that work groups initiated debiasing efforts and became aware of biases. Debiasing then however lost its significance in favor of quick analysis results and decision-making. More awareness of the pitfalls of biases might shift the focus to debiasing first, before final information products are developed and decisions are made. Post-mortem analysis of information management and decision-making processes after crisis response can be implemented in lesson learnt and debriefing sessions. Further, large-scale crisis response trainings, which are organized annually by major response organizations to train together for real crisis event (e.g., SIMEX, TRIPLEX), should incorporate debias interventions in training agendas.

Second, the development of models, algorithms and information systems to support information management and decision-making in crisis response, should implement functions that help identify and mitigate biases in (a) the datasets used by these systems, and (b) the cognitive processes of system users.

\section{Limitations}
In our paper, we present an initial exploratory study on the interplay of data and confirmation bias in time-critical sequential decisions. Because of the exploratory nature of our study, there are several limitations that can be addressed in future research. 

First, and to the best of our knowledge, while our study is the first of its kind that brings external analysts together with decision-makers to study their joint CIM process in a realistic scenario-based experiment, and our participants were all experienced in their roles, the number of participants is a limiting factor in our study. Similar studies have reported larger participant groups, mostly of inexperienced students and other laypersons who are easy to recruit. We suggest to expand on our findings in additional larger-scale experiments and surveys across diverse groups and different professional experiences.  

Our experimental design was inspired by hidden profile experiments. In traditional hidden profile experiments \citep{Stasser1985, Lightle2009}, participants are asked to study their received information before joining the group conversation. In contrast, we allowed for discussions from the start because crisis information management is characterized by fast, agile communication. Our approach decreased the chances that participants constructed a rigid mental model of what data they received initially. Two characteristics of our research design counter this shortcoming. First, we allowed for perfect recall, i.e., participants kept all materials during the workshop experiment. Second, participants needed to continuously engage with the data by aggregating, analyzing, and visualizing it, so they had to build a deep understanding of the data during the experiment.

It is a major challenge to simulate a realistic crisis environment in an experimental setting. This includes a realistic but still unknown scenario, decision-making under urgency, uncertainty, high stakes, and constraint resources, allowing for interactive collaboration with multiple actors, and providing equipment that resembles experts' real work environment. Simplifications have to be made to make the experiments controllable. In addition, we had to consider that some organizations might implement and pursue different approaches to information management and decision support than required by the tasks we set. In real-world scenarios, external analysts work with a larger group of colleagues. Because of the framework required by our experiment, for example, the discussions on the creation of the information products had to be objectively observed on site, it was not possible to include further external analysts from those remotely working communities. Here, we suggest to complement our findings with more ethnographic and field studies in real disaster to observe real-world debiasing and decision-making behaviour.

\section{Conclusion}
\label{sec:conclusion}

Crisis response organizations integrate external analysts into the CIM process to strengthen their digital resilience. In this capacity, external analysts collect and analyze data and develop information products (e.g., maps, tables, infographics) for decision support. While this extended capacity is meant to improve the evidence base for decisions, the CIM process remains challenged by circumstances of urgency, uncertainty, high stakes, and constraint resources. Consequently, crises are prone to induce biases into the data as well as the cognitive processes of external analysts and decision-makers. We investigated how biases influence the CIM process between experienced external analysts and decision-makers through a three-stage experiment. 

Our findings show that data biases, even if detected, influence the development of information products for crisis decision support. We show that effective debiasing does not happen because crisis information managers have a strong commitment and urgency to deliver a presentable information product that is actionable enough for decision-makers to make decisions directly. Efforts for creating information products are prioritized, and debiasing is neglected. In subsequent deliberations and decision-making discussions, decision-makers are influenced by biased information products in their allocation decisions of scarce resources. Confirmation bias amplifies the reliance on problematic assumptions that were made based on biased data. This implies that the biased, misleading information that shapes initial decisions is perpetuated by a vicious circle of biased information search that influences future decisions. Our findings indicate that decisions in crisis response can only be effective if initial data and confirmation bias are identified and mitigated. Mindful debiasing could be a successful strategy to improve broad information search and tackle both biases.

\section{Acknowledgement}
\label{sec:acknowledgement}

This work was funded through the Special Priority Program “Volunteered Geographic Information: Interpretation, visualization and Social Computing” (SPP 1894) by German Research Foundation (DFG, Project number: 273827070). We thank all participants and especially the student assistants who supported the preparation and organization of the experiments and observations during the sessions.


%
%


\section{Conflict of interests}
The authors have no conflicts of interests to declare.

\bibliography{bibtexlib}

\pagebreak

\begin{appendices}
\section{- Observation protocol}

\begin{table}[H]
\resizebox{\textwidth}{!}{%
\begin{tabular}{|l|l|l|}
\hline
\textbf{A. General description on site}                                                                                                                                                              & \textbf{\begin{tabular}[c]{@{}l@{}}B. Communication and interaction \\ description\end{tabular}}                                                                                                             & \textbf{C. General impressions}                                                                                     \\ \hline
\begin{tabular}[c]{@{}l@{}}How does the workspace you\\  are observing look? (Seating \\ arrangement, communication \\ devices, support materials, \\ additional characteristics, etc.)\end{tabular} & \begin{tabular}[c]{@{}l@{}}Describe the sequence of \\ events over time (e.g., information\\  search, prioritization, processing, \\ request, sharing, group discussion, \\ decision-making, …)\end{tabular} & \begin{tabular}[c]{@{}l@{}}Tone of the discussion (rational, \\ empathic, humorous, etc.)\end{tabular}              \\ \hline
Participant coding                                                                                                                                                                                   & \begin{tabular}[c]{@{}l@{}}Which information is shared \\ among the participating V\&TCs?\end{tabular}                                                                                                       & Speedy vs. lengthy discussions?                                                                                     \\ \hline
\begin{tabular}[c]{@{}l@{}}Was communication rather\\  face-to-face or mediated \\ via technology?\end{tabular}                                                                                      & \begin{tabular}[c]{@{}l@{}}Are additional information sources \\ used?\end{tabular}                                                                                                                          & \begin{tabular}[c]{@{}l@{}}Attitude of individual participants \\ (engaging, negative, overwhelmed, …)\end{tabular} \\ \hline
                                                                                                                                                                                                     & \begin{tabular}[c]{@{}l@{}}How is the need for information\\  expressed and communicated?\end{tabular}                                                                                                       & \begin{tabular}[c]{@{}l@{}}To what extent was available\\  information not shared / retained?\end{tabular}          \\ \hline
                                                                                                                                                                                                     & \begin{tabular}[c]{@{}l@{}}Which decisions are anticipated \\ to be supported by the V\&TCs?\end{tabular}                                                                                                    & Additional comments                                                                                                 \\ \hline
                                                                                                                                                                                                     & \begin{tabular}[c]{@{}l@{}}Describe how and why specific\\  types of information products are \\ selected and created for the \\ decision-makers.\end{tabular}                                               &                                                                                                                     \\ \hline
                                                                                                                                                                                                     & \begin{tabular}[c]{@{}l@{}}Which information is included\\  and why?\end{tabular}                                                                                                                            &                                                                                                                     \\ \hline
                                                                                                                                                                                                     & \begin{tabular}[c]{@{}l@{}}Which technology and other \\ decision aid materials are \\ utilized and how?\end{tabular}                                                                                        &                                                                                                                     \\ \hline
\end{tabular}%
}
\end{table}

\section{- Confirmation bias measure}
“Below are the summaries of 10 new datasets that are available. You can re-quest the full version of those datasets but you only have limited time and re-sources to evaluate them all in detail. Select as many datasets as you want. District X is the district you have identified in the last session as the most critical district.”
\begin{itemize}
\item Dataset 1: District X has less treatment capacity than infection cases.
\item Dataset 2: In district X the infection rate is likely to increase.
\item Dataset 3: District X has high infrastructural damage.
\item Dataset 4: District X has a low percentage of people reached.
\item Dataset 5: District X has more treatment capacity than infection cases.
\item Dataset 6: In district X the infection rate is likely to decrease.
\item Dataset 7: District X has low infrastructural damage.
\item Dataset 8: District X has a high percentage of people reached.
\item Dataset 9: District X has a high amount of health care workers infected.
\item Dataset 10: District X has a low amount of heath care workers infected.
\end{itemize}

\section{- Author biographies}
David Paulus is a PhD researcher at the Faculty of Technology, Policy and Management at Delft University of Technology. He studies data biases and cognitive biases in humanitarian information management and decision-making. He is Delft Global Fellow since 2019 and member of the International Association for Information Systems in Crisis Response and Management (ISCRAM) since 2016. In his research he combines theories and methods from computer science, psychology and organizational science. As a Research Associate at the United Nations University Institute for Environment and Human Security from 2015-2017, he was involved in ICT-supported institutional capacity building projects in North Africa and Southeast Asia.

Ramian Fathi is a research associate and PhD candidate at the Chair for Public Safety and Emergency Management, University of Wuppertal. As part of the DFG-funded Priority Programme (1894) “Volunteered Geographic Information”, his research focusses on the analysis of social media by digital volunteers and their participation in disaster management. In addition, he is the team leader of the Virtual Operations Support Team (VOST) of German Federal Agency for Technical Relief (THW) and vice president of the German Society for the Support of Social Media and Technology in Civil Protection (DGSMTech e.V.).

Dr. Frank Fiedrich holds the Chair for Public Safety and Emergency Management at the University of Wuppertal since 2009. He studied Industrial Engineering and received his Ph.D. from the Karlsruhe Institute of Technology, Germany, where he worked on Decision Support Systems and Agent-based Simulation for disaster response. From 2005 to 2009, he was Assistant Professor at the Institute for Crisis, Disaster, and Risk Management ICDRM at the George Washington University, Washington DC. His research interests include the use of information and communication technology for disaster and crisis management, societal, organizational and urban resilience, interorganizational decision-making, critical infrastructure protection and societal aspects of safety and security technologies. Additionally, Professor Fiedrich is honorary member of the International Association for Information Systems in Crisis Response and Management (ISCRAM).

Dr. Bartel Van de Walle is a UN diplomat and director of the United Nations University Institute UNU-MERIT in Maastricht, the Netherlands. UNU-MERIT carries out research and training on a range of social, political and economic factors that drive economic development in a global perspective. Dr. Van de Walle is also professor of policy analysis for global challenges at Maastricht University. His research focuses on humanitarian response, and specifically on the role of information systems for better coordination and response. He is a member of the steering committee for the Dutch Science Foundation's Science for Global Development initiative.

Dr. Tina Comes is Full Professor on Decision Theory and Information Technology for Resilience  at the TU Delft, Netherlands, and Full Professor in Decision-Making and Digitalisation at the University of Maastricht. Dr. Comes is a Visiting Professor at the Université Dauphine, France, a member of the Norwegian Academy for Technological Sciences and the Academia Europaea. She serves as the Scientific Director of the 4TU.Centre for Resilience Engineering, as Principal Investigator on Climate Resilience for AMS, as Director of the TPM Resilience Lab, and she leads the Disaster Resilience theme for the Delft Global Initiative. Prof. Comes’ research focuses on decision-making and information technology for resilience and disaster management. This perspective on decision making, resilience and humanitarian response is reflected in more than 100 publications.

\end{appendices}

\end{document}